\documentclass[twocolumn,english]{asme2e}
\usepackage[T1]{fontenc}
\usepackage[latin9]{inputenc}
\usepackage{float}
\usepackage{amsmath}
\usepackage{amssymb}
\usepackage{graphicx}

\makeatletter

\providecommand{\tabularnewline}{\\}
\floatstyle{ruled}
\newfloat{algorithm}{tbp}{loa}
\providecommand{\algorithmname}{Algorithm}
\floatname{algorithm}{\protect\algorithmname}

\usepackage{tabularx}
\usepackage{array}
\providecommand{\tabularnewline}{\\}

\usepackage{graphicx}
\iftrue
        \newcommand{\cutsectionup}{\vspace*{-0.1in}}

        \newcommand{\cutsubsectionup}{\vspace*{-0.09in}}

        \newcommand{\cutequationup}{\vspace*{-0.12in}}
        \newcommand{\cutequationdown}{\vspace*{-0.12in}}

\else % do not use space-saving macro
        \newcommand{\cutsectionup}{}

        \newcommand{\cutsubsectionup}{}

        \newcommand{\cutequationup}{}
        \newcommand{\cutequationdown}{}

\fi

\conffullname{Working Paper}
\confmonth{March}
\confyear{2018}

\usepackage{babel}

\makeatother

\usepackage{babel}
\begin{document}

\title{PREDICTING ``DESIGN GAPS\textquotedbl{} IN THE MARKET: DEEP CONSUMER CHOICE MODELS UNDER PROBABILISTIC DESIGN CONSTRAINTS}

\author{Alex Burnap\thanks{Address all correspondence to this author.}\affiliation{%Sloan
%School of Management\\
 Massachusetts Institute of Technology\\
 Cambridge, MA 02139\\
 Email: aburnap@mit.edu}}

\author{John R. Hauser\affiliation{%Sloan School of Management\\
 Massachusetts Institute of Technology\\
 Cambridge, MA 02139\\
 Email: hauser@mit.edu}}
\maketitle
\begin{abstract}
Predicting future successful designs and corresponding market opportunity
is a fundamental goal of product design firms. There is accordingly
a long history of quantitative approaches that aim to capture diverse
consumer preferences, and then translate those preferences to corresponding ``design gaps\textquotedblright{} in the market. We extend this work by developing a deep learning approach to predict design gaps in the market. These design gaps represent
clusters of designs that do not yet exist, but are predicted to be both (1) highly preferred by consumers, and (2) feasible to build under engineering and manufacturing
constraints. This approach is tested on the entire U.S. automotive market
using of millions of real purchase data. We retroactively predict
design gaps in the market, and compare predicted design
gaps with actual known successful designs. Our preliminary results give evidence
it may be possible to predict design gaps, suggesting this approach
has promise for early identification of market opportunity.
\end{abstract}
% \cutsectionup
% \cutsectionup
% \cutsectionup
\cutsectionup 

\section{Introduction}

\label{sec:Introduction}

Predicting consumer preferences for future design concepts is a fundamental
task for product design enterprises \cite{krishnan2001product}. For
example, consumer preferences drove ``crossover vehicles'' to overtake
the previously dominant SUV segment within just years of introduction
to the U.S. automotive market \cite{korenok2010nonprice}.

To predict consumer preferences, quantitative models of human choice
have been researched and developed for more than a century \cite{thurstone1927themethod,tverskyelimination}.
In general, these methods estimate a predictive model of future consumer
preferences using stated or revealed data from past consumer choices
over design alternatives. These predictive models can then be used
to improve design or managerial decisions such as predicting consumer
demand \cite{rossi2003bayesian} or market segmentation \cite{wedel2012marketsegmentation}.
The success of these methods is underscored by their adoption by product
design enterprises\textendash one of the most popular methods, conjoint
analysis, has over 18,000 applications a year \cite{orme2010getting}.

One of the most challenging consumer preference prediction tasks is
to identify ``design gaps\textquotedbl{} in the market. These design
gaps are clusters of designs that do not yet exist, are perhaps unknown
to the firm, yet represent potentially significant market opportunity.
Early identification of design gaps offers firms competitive
advantage due to first mover effects.

The challenge in predicting design gaps is due to the number of statistical
unknowns. One can conceptually group consumer preference prediction
tasks by their unknowns into three categories of increasing difficulty:
The first category, with the least unknowns, is to predict what \textit{unknown}
consumers would prefer among \textit{known and existing} product designs;
e.g., product recommendation by online retailers. The second category
is to predict what \textit{unknown} consumers would prefer among \textit{known
and not existing} product designs; e.g., SUV concept designs during
prototyping. The third category predicts what \textit{unknown}
consumers would prefer among \textit{unknown and not existing} product
designs.

In this work, we address this third category. This aims at the question:
Can firms quantitatively predict design gaps that represent potential
market opportunities at the earliest stages? With new product development,
could automotive manufacturers have predicted years in advance that
the crossover vehicle segment represented a massive market opportunity?
With updating existing products, could the same manufacturers identify
new combinations of infotainment features and marketing cues that
appeal to millenials?

We introduce a deep learning approach that aims to identify design
gaps hidden in large-scale data. This approach predicts regions of the design space that (1)
have high predicted consumer preference, and (2) are feasible to build
according to probabilistic estimates of design constraints from engineering,
manufacturing, and other downstream design processes. This approach
combines and builds on research from quantitative marketing in consumer
choice modeling, and engineering design for bounding design feasibility
within the design space.

We test this approach on several years of millions of actual purchase
data from the U.S. automotive market. To artificially induce \textquotedbl{}design
gaps,\textquotedbl{} we use actual purchase data to retroactively
construct \textit{unknown} consumers and \textit{unknown and not existing}
product designs. Validity is then assessed by how well the proposed
deep learning approach predicts design space regions containing held-out
new design entrants to the market.

Our preliminary results give evidence that this approach may lead
to design gaps. This offers the following contributions:
\begin{enumerate}
\item Deep heterogeneous consumer choice models can significantly improve
held-out choice prediction accuracy.
\item Deep unsupervised models can efficiently estimate probabilistic design
constraints to bound the design space. 
\item Given (1) and (2), prediction of design gaps may be possible using
information-theoretic disaggregate choice metrics. 
\end{enumerate}
This approach does not predict the actual design itself, nor does
it directly predict market opportunity. It instead aims to highlight
significantly reduced promising subsets of the feasible design space
rather than the otherwise exponential number of possible designs.
In other words, this approach still requires designers and market
researchers investigate proposed design gaps, offering instead a scalable
data-driven approach to augment both strategic marketing and the creative
design process.

The rest of this paper is structured as follows: Section \ref{sec:Related-Work}
discusses related work in marketing and design. Section \ref{sec:Problem-Formulation}
formulates a consumer choince model under design constraints. Section
\ref{sec:Experiment} conducts an experiment using real purchase data.
Section \ref{sec:Discussion} discusses implications and opportunities
for future work. Section \ref{sec:Conclusion} offers conclusions.

\begin{figure}
\begin{centering}
\includegraphics[width=1\linewidth]{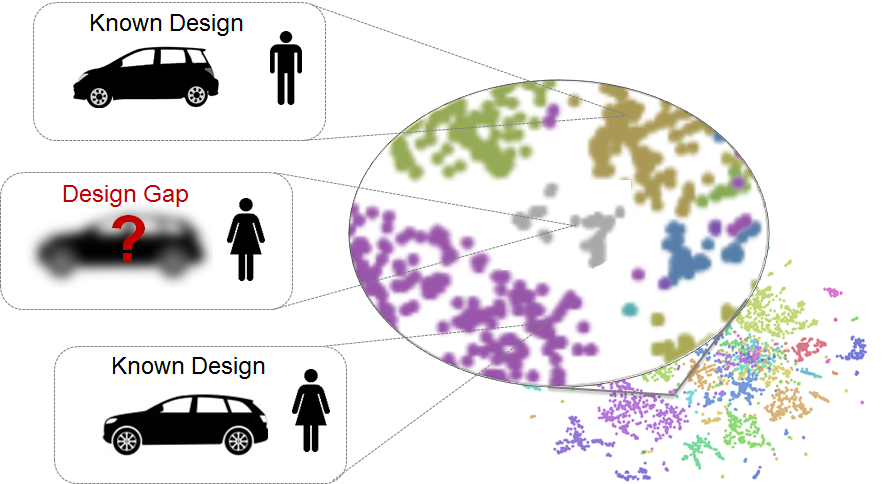}
\par\end{centering}
\caption{High-level goal of this work, predicting potentially unknown ``design
gaps'' in the market.\label{fig:design-gap-into}}
\end{figure}

\section{Related Work}

\label{sec:Related-Work}We describe related work primarily in two
communities, quantitative marketing and engineering design. Our work
 builds on conventions in each of these communities.

\cutsubsectionup
\subsection{Consumer Preference Modeling}

There is long and rich history of modeling techniques aimed at capturing
consumer preference and choice decisions over a set of alternatives,
across fields such as psychology, economics, marketing, and product
design. 

Most relevant to our work is the research carried out in capturing
consumer choice in a probabilistic model used to inform design decisions.
One of the most popular models is this group is conjoint analysis,
in which consumers have a ``part-worth vector'' describing their
affinity for design attributes (e.g., color, price, horsepower) through
a linear relationship. These models have their roots in psychology
\cite{thurstone1927themethod,tversky1969intransitivity,luce1959individual}
and economics \cite{vonneumann2007theoryof}, but received major attention
for improving design and marketing during the supermarket scanner
data revolution \cite{rossi1996thevalue}.

Since this first data revolution, several extension to capture the diversity of consumer
preferences, referred to as consumer heterogeneity, have received
considerable attention. An early yet still widely used model of heterogeneity
extends the basic conjoint formulation to having a hierarchical prior distribution \cite{rossi2003bayesian}. This prior encodes the idea that the population
is centered around an average consumer, and that heterogeneity is
related to the distance from the average. In general, the notion of modeling consumer heterogeneity improves tasks such as consumer demand or market segmentation
\cite{wedel2012marketsegmentation}.

Within design research, consumer choice models are found
widely in decision-based formulations of design \cite{hazelrigg1998aframework,wassenaar2005enhancing,lewis2006decision}.
Recently several works have focused on modeling heterogeneous consumer consideration
for designs \cite{morrow2014marketsystem}, building off again original
findings in psychology \cite{payne1976taskcomplexity}. These formulations
note that preference decisions are not made randomly over set of existing
designs, but rather a subset where attributes are traded off \cite{hauser2010disjunctions}.

Another key concept adopted by our work is the idea of higher-level
abstractions of the preference task. For example, consumers do not
necessarily purchase a vehicle for its number of valves per cylinder, but rather
meaningful attributes such as perceived luxuriousness and external image. This
notion has been observed in many guises across design \cite{norman2003affectand,norman2004emotional},
psychology \cite{brunswik1952conceptual}, and marketing \cite{tybout1981amarketing}.

Our work aims to estimate both consumer heterogeneity and  higher-level preference abstractions as ``features''
discovered from the data \cite{burnap2016improving}. We place flexible mathematical assumptions on the form of consumer heterogeneity,
instead letting it be ``learned\textquotedblright{} by recent advances
in deep learning over large data sets. Our work is similar in this
guise to machine learning work in marketing \cite{toubia2001fastpolyhedral} for efficient consumer choice estimation
algorithms, as well as flexible function forms on preference itself
\cite{evgeniou2005generalized, liu2016capturing}. In addition, we use the well-known concept of ``borrowing strength''
of Bayesian conjoint analysis models \cite{lenk1996hierarchical}, albeit with a deep neural network
formulation.

\cutsubsectionup
\cutsubsectionup
\subsection{Design Constraint Modeling}

The aim of modeling design feasibility is to capture constraints on
new potential design concepts, physical or perhaps implicit from some
other source. For example, the design may have engineering constraints
due to physical relations between aerodynamics and vehicle handling dynamics, physical manufacturing
constraints based on scalability of existing powertrain or chassis
platforms, and perceptual constraints such as brand recognition and
perceived image cues \cite{burnap2016balancing}.

These constraints are important when identifying design gaps. Specifically, one should take reservation when optimizing consumer preferences with respect to an
unbounded space of designs. Such an optimization can lead to the so-called ``million-dollar car," which theoretically has high consumer demand but is unrealistic due to violations of engineering and fiscal constraints.

Several major pathways towards design representations that encode realistic constraints have been developed by the design community. Direct analytic modeling uses deterministic
mathematical relationships amongst variables. Conceptually related to our work is \cite{michaleklinking}, who combine consumer choice
models and analytic design models. Recent work includes mixing sophisticated consumer choice consideration models with analytic design models \cite{long2015shouldoptimal}. Simulation-based approaches to design feasibility
are widely-used, for example,  \cite{bayrak2013designof}, who develop an explicit generative model of all hybrid vehicle powertrain architectures.
Their model has conceptual similarities with our design gap identification approach, as both perform rejection sampling. Rule-based design representations
use smaller elements and rules on design synthesis. For example, \cite{singh2012towards}
and \cite{perezmata2015spatial} use shape grammars with aesthetic
rules for a parametrized vase designs.

Most related to our work is that of using probabilistic models of
design representations. As noted in \cite{vanhorn2012designanalytics},
large-scale data is enabling better capture of consumer choice preferences
for early-stage design. \cite{tuarob2015aproduct} collect large-scale
social media data for mining consumer preferences. Similar to our
work is \cite{osogami2014restricted}, who use restricted Boltzmann
machines to model consumer choices. Likewise, in marketing, \cite{hruschka2014analyzing} use the same restricted Boltzmann machine for predicting market baskets,
and note its relationship with multinomial logit models. Our work
is similar in that we use a multinomial logit model with a ``deep''
architecture for both learning consumer preferences, as well as capturing design constraints.  As a result, our approach is a combination of both discriminatative and generative models \cite{mnih2007probabilistic,salakhutdinov2007restricted,osogami2014restricted}.

\cutsubsectionup
\cutsectionup
\cutsectionup
\section{Problem Formulation\label{sec:Problem-Formulation}}

The high-level goal of the following mathematical formulation is to
capture the relationship between consumers and product designs that are realistically feasible. In
particular, we are interested in an accurate relationship between
not just consumers and designs that currently exist, but between consumers
and designs that are theoretically possible. This level of generalization
accuracy requires not just capturing the relationship between consumers
and designs, but capturing the relationship between existing designs
and future designs as well as existing consumers and future consumers.

Let us denote consumers $\mathbf{x}_{c}\in\mathcal{X}_{c}\subseteq\mathbb{R}^{M_{c}}$
and designs $\mathbf{x}_{d}\in\mathcal{X}_{d}\subseteq\mathbb{R}^{M_{d}}$
using vectors, and denote the $c^{th}$ customer as $\mathbf{x}_{c}^{(c)}$
and the $d^{th}$ existing design as $\mathbf{x}_{d}^{(d)}$. The
consumer space $\mathcal{X}_{c}$ and design space $\mathcal{X}_{d}$
are both bounded subsets of the real values. We also assume that there
is a finite set of existing design alternatives for consumers to choose
from $\{\mathbf{x}_{d}^{(d)}\}_{d=1}^{D}$, in which the true purchased
or preferred design is denoted as an indicator vector $\mathbf{y}\in\{0,1\}^{D}$
over all existing designs, as well as the shorthand $y^{(d)}\in\{0,1\}$ as
the $d^{th}$ element of the indicator vector $\mathbf{y}$.

Quantitative choice models assume a parametric utility function $U(\mathbf{x}_{c},\mathbf{x}_{d},\theta):\mathcal{X}_{c}\times\mathcal{X}_{d}\rightarrow\mathbb{R}$,
where $\theta$ are parameters of the utility model that index consumer
and design pairs to a real number value in which higher numbers correspond
to higher preference \cite{vonneumann2007theoryof}. This utility
function lies within a probabilistic choice or preference model, in
which the choice model maps the consumer response task (e.g., ranking,
rating, choice) to values of utility \cite{luce1959individual}. In
doing so, the choice model accounts for uncertainty from factors
such as missing latent variables, model mispecification, and measurement
error \cite{rossi2003bayesian}.

We aim to predict how much the $c^{th}$ consumer $\mathbf{x}_{c}^{(c)}$
prefers the $d^{th}$ product design $\mathbf{x}_{d}^{(d)}$ among
the entire set of $D$ designs. This invokes representing the $c^{th}$
consumer's utility function for the $d^{th}$ design in relation to
all other designs $i=1\ldots D$. 

\cutequationup
\cutequationup
\cutequationup
\begin{align}
 & p(y^{(d)}=1|\mathbf{x}_{c},\mathbf{x}_{d})\label{eq:multinomial-model-original}\\
 & =p\left(U\left(\mathbf{x}_{c}^{(c)},\mathbf{x}_{d}^{(d)}\right)+\epsilon^{(c,d)}>\left\{ U\left(\mathbf{x}_{c}^{(c)},\mathbf{x}_{d}^{(i)}\right)+\epsilon^{(c,i)}\right\} _{i=1}^{D}\right)\nonumber 
\end{align}
where $\epsilon^{(c,d)}$ is extreme value distributed \cite{gumbel1958statistics},
resulting in a softmax function for choice probability \cite{luce1959individual}.

The full Bayesian joint likelihood $\mathcal{L}_{A}(\theta;\mathbf{y},\mathbf{x}_{c},\mathbf{x}_{d})$ of $p$ over $\mathcal{X}_{c}\times\mathcal{X}_{d}\times\mathcal{Y}$
acts as a generative distribution of consumer preference under design
constraints that bound $\mathcal{X}_{d}$ and consumer heterogeneity
that bound $\mathcal{X}_{c}$. We take a fully Bayesian approach, in that we invoke prior distributions over
consumers $p(\mathbf{x}_{d})$ and design $p(\mathbf{x}_{d})$.

\cutequationup
\begin{eqnarray}
\mathcal{L}_{A}(\theta;\mathbf{y},\mathbf{x}_{c},\mathbf{x}_{d}) & = & p(\mathbf{y}|\mathbf{x}_{c},\mathbf{x}_{d})p(\mathbf{x}_{d})p(\mathbf{x}_{c})\nonumber \\
p(\mathbf{y}|\mathbf{x}_{c},\mathbf{x}_{d}) & = & \mathrm{M}(\mathbf{y})\nonumber \\
p(\mathbf{x}_{d}) & = & \mathcal{N}\left(\mathbf{x}_{dr};\boldsymbol{\mu},\Sigma_{d}\right)\prod_{b}\mathrm{B}(x_{db,b})\prod_{m}\mathrm{M}(\mathbf{x}_{dm,j})\label{eq:bayesian-model-original}\\
p(\mathbf{x}_{c}) & = & \frac{1}{N}\sum_{i}\delta(\mathbf{x}_{c}-\mathbf{x}_{c}^{(i)})\nonumber 
\end{eqnarray}
where $\mathcal{N}\left(\cdot\right)$ is the multivariate Gaussian
density with mean vector $\boldsymbol{\mu}$ and covariance matrix
$\Sigma_{d}$, $\mathrm{B}(\cdot)$ is the Bernoulli density with
the sigmoid function $p(x_{db,b}=1)=\frac{e^{\theta_{b}x_{d}}}{1+e^{\theta_{b}x_{d}}}$
, $\mathrm{M}(\cdot)$ is the multinomial density using the softmax
function $p(\mathbf{y}|\mathbf{x}_{c},\mathbf{x}_{d})=\frac{e^{U\left(\mathbf{x}_{c},\mathbf{x}_{d}^{(d)}\right)}}{\sum_{i}e^{U\left(\mathbf{x}_{c},\mathbf{x}_{d}^{(i)}\right)}}$
for consumer choice preferences or $p(\mathbf{x}_{d,m}=k)=\frac{e^{\theta_{m}^{(k)}\mathbf{x}_{d}}}{\sum_{i}e^{\theta_{m}^{(i)}\mathbf{x}_{d}}}$
for categorical design variables, and $\delta(\cdot)$ is a Dirac
delta function used to represent the empirical distribution of consumers.

\subsection{Feature Learning}

The consumer choice prediction model given in Equation (\ref{eq:multinomial-model-original})
has been widely used across disciplines such as psychology, marketing,
economics, and design. While the underlying mathematical
formulation is conceptually the same, common modeling challenges are
often addressed in separate manners according to the conventions of
the field. This offers us the opportunity to build on modeling assumptions
across fields, making explicit our modeling goals in an effort to
improve predictive model fidelity.

\subsubsection{Consumer Heterogeneity and Design Feasibility}

One of the major challenges in the choice modeling research is how to
represent consumer preference heterogeneity; in short, the diversity
in human preferences. Mathematical representations to encode heterogeneity
involve assuming a distribution over customer variables as they relate
to the design. Implicitly, this distribution encodes constraints on
the space of consumers $\mathcal{X}_{c}$; for example, certain hobbies
may be constrained by income and leisure time.

Quantitative marketing methods often explicitly encode consumer constraints,
such as monotonic relationships on price sensitivity \cite{orme2010getting}. Ever more sophisticated
constraint mechanisms often integrate research marketing and psychology
research findings, such as partitioning the set of existing designs
$\{\mathbf{x}_{d}^{(d)}\}_{d=1}^{D}$ into a much smaller consideration
set \cite{hauser2010disjunctions,payne1976taskcomplexity,morrow2014marketsystem},
or violations of rationality assumptions \cite{kahneman2003mapsof}.

At the same time, one of the major challenges in engineering design
is the formulation of design representations that are flexible, yet
physically accurate. As described in Section \ref{sec:Related-Work},
design representations include several formulations. Common to all
of them is the desire to encode constraints for the space of designs
$\mathcal{X}_{d}$, in which the diversity of possible designs is
captured.

This analogy between constraints on the diversity of consumer heterogeneity
and the diversity of design constraints allows us to take a similar
probabilistic approach between these two spaces. Our approach uses
large-scale data to learn the set of constraints that bound both consumers
and designs.

\subsubsection{Feature Representation}

To estimate the diverse constraints in both the consumer space and
design space, we build on substantial research that consumers consider
``attributes'' or ``features'' of the design (e.g., `perceived
luxuriousness') rather than the original variables (e.g., `valves
per cylinder'). We aim to learn consumer features $\mathbf{h}_{c}\in\mathcal{H}\in\mathbb{R}^{K}$
and design features $\mathbf{h}_{d}\in\mathcal{H}\in\mathbb{R}^{K}$ that efficiently
represent and constraint the information about the consumer variables $\mathbf{x}_{c}$
and design variables $\mathbf{x}_{d}$ at a more abstract level closer
to the consumer's actual preference task.

Furthermore, we aim to impose ``useful\textquotedbl{} probabilistic
structure for this new feature representation of the design space
$p(\mathbf{h}_{d}|\mathbf{x}_{d})p(\mathbf{x}_{d})$, while keeping
$\mathbf{h}_{d}$ maximally informative about the consumer preference
choice $\mathbf{y}$. This structure is important for design gap prediction
as it allows us a manageable representation of the otherwise high-dimensional,
highly-nonlinear, and discontinuous structure of the original variables
$\mathbf{x}_{d}$ of the design space (e.g., interpolations
between $\frac{1}{2}$-ton trucks and turbocharged convertibles).
Specifically, learning a ``smoother'' space allows reasonable
and efficient sampling and optimization, while dimensionality reduction
helps ameliorate the otherwise exponential number of possible concept
design.

Accordingly, the original choice model given in Equation (\ref{eq:multinomial-model-original})
is now updated with a new utility formulation $U(\mathbf{h}_{c},\mathbf{h}_{d},\theta):\mathcal{H}^2\rightarrow\mathbb{R}$ as follows.

\begin{align}
p\left(U\left(\mathbf{h}_{c}^{(c)},\mathbf{h}_{d}^{(d)}\right)+\epsilon^{(c,d)}>\left\{ U\left(\mathbf{h}_{c}^{(c)},\mathbf{h}_{d}^{(i)}\right)+\epsilon^{(c,i)}\right\} _{i=1}^{D}\right)\label{eq:feature-choice-model}
\end{align}

We assume conditional independencies between preference responses
and the original variables given on their feature representations.
This assumption ensures customer and design features contain the maximal
possible information for the consumer preference prediction task.
Our equivalent learning objective is now to predict how much the $c^{th}$
consumer $\mathbf{h}_{c}^{(c)}$ prefers the $d^{th}$ product design
$\mathbf{h}_{d}^{(d)}$ among the entire set of $D$ designs. This
updates Equations (\ref{eq:multinomial-model-original}) and (\ref{eq:bayesian-model-original})
by representing the $c^{th}$ consumer's utility function for the
$d^{th}$ design in relation to all other designs $i=1\ldots D$. 

The full Bayesian joint probability distribution $p$ over $\mathcal{X}_{c}\times\mathcal{X}_{d}\times\mathcal{Y}$
is now,

\cutequationup
\cutequationup
\begin{eqnarray}
 & \mathcal{L}_{A}(\theta;\mathbf{y},\mathbf{x}_{c},\mathbf{x}_{d}) & =p(\mathbf{y}|\mathbf{x}_{c},\mathbf{x}_{d})p(\mathbf{x}_{d})p(\mathbf{x}_{c})\\
 &  & =p(\mathbf{y}|\mathbf{h}_{c},\mathbf{h}_{d})p(\mathbf{h}_{c}|\mathbf{x}_{c})p(\mathbf{h}_{d}|\mathbf{x}_{d})p(\mathbf{x}_{d})p(\mathbf{x}_{c})\nonumber
\end{eqnarray}
\cutequationdown
\cutequationdown
\cutequationdown

% These feature representations encode physical and implicit constraints
% of both the design (e.g., `fuel efficiency tradeoff with powertrain'),
% as well as the consumer (e.g., `hobbies relation with with income'),
% purely from the data. 
Learning these feature representations is challenging due to both the heterogeneous variable types of the
data (i.e., real, binary, categorical), as well as the desire to encode ``useful'' probabilistic structure
in the data. Note that the inner product between consumers $\mathbf{h}_{c}$
and designs $\mathbf{h}_{d}$ ensure that consumer and design information
interacts (i.e., we are not just performing classification). In practice,
we estimate $p(\mathbf{y}|\mathbf{h}_{c}\mathbf{h}_{d})=\mathrm{M}(\mathbf{y}|\mathbf{h}_{c}^{T}\mathbf{H}_{d})$
, where $\mathbf{H}_{d}$ is a $K\times D$ matrix of all existing
designs.

\cutsubsectionup
\cutsubsectionup
\subsection{Deep Choice Model under Design Constraints}

To impose the ``useful\textquotedbl{} probabilistic structure required
to efficiently traverse the feature representation of the design space
$p(\mathbf{h}_{d}|\mathbf{x}_{d})p(\mathbf{x}_{d})$, we now introduce
modeling assumptions that aim to impose structure while being amenable
to recent advances in deep learning. 

To this end, we adopt an approach recently popularized in deep learning
called black-box variational inference. This approach aims to approximate
complex probability distributions by variationally bounding the complex
distribution with a more manageable joint distribution with introduced
latent random variables \cite{braun_variational_2010}. In particular, we build on variational autoencoders
as introduced in \cite{kingma2013autoencoding}.

Our approach makes two minor extensions. First, given that we are
working in a supervised learning regime (i.e., we eventually predict
consumer preferences over new concept designs to discover ``design
gaps''), we include an extension that introduces an information signal
to the design feature representation from the preference task. This extension is conceptually similar to semi-supervised approaches with a known labels for the encoded design representation, but different in that our supervision occurs after interaction with the consumer space \cite{kingma_semi-supervised_2014,siddharth_learning_2017}. Second,
given the heterogeneity of variable types for our data (i.e., real,
binary, and categorical), we include exponential family derivations
of the corresponding likelihoods in the original design variable space.

Taking the original likelihood function $\mathcal{L}_{A}(\theta,\phi;\mathbf{y},\mathbf{x}_{c},\mathbf{x}_{d})$, we introduce a approximation
density $q(\mathbf{h}_{d}|\mathbf{x}_{d})$, which we will use to invoke desired ``useful'' probabilistic structure
for the design feature representation $\mathbf{h}_{d}$.

\cutequationup
\begin{eqnarray}
 & \mathcal{L}_{A}(\theta,\phi;\mathbf{y},\mathbf{x}_{c},\mathbf{x}_{d}) & =\int_{\mathbf{h}_{d}}\log[p(\mathbf{y},\mathbf{x}_{c},\mathbf{x}_{d})]q(\mathbf{h}_{d}|\mathbf{x}_{d})d\mathbf{h}_{d}\\
 &  & =\int_{\mathbf{h}_{d}}q(\mathbf{h}_{d}|\mathbf{x}_{d})[\log\frac{p(\mathbf{y},\mathbf{x}_{c},\mathbf{h}_{d},\mathbf{x}_{d})}{q(\mathbf{h}_{d}|\mathbf{x}_{d})}\nonumber \\
 &  & +\log\frac{q(\mathbf{h}_{d}|\mathbf{x}_{d})}{p(\mathbf{h}_{d}|\mathbf{y},\mathbf{x}_{d})}]d\mathbf{h}_{d}\nonumber \\
 &  & =\mathbb{E}_{q(\mathbf{h}_{d}|\mathbf{x}_{d})}\left[\log\frac{p(\mathbf{y},\mathbf{x}_{c},\mathbf{h}_{d},\mathbf{x}_{d})}{q(\mathbf{h}_{d}|\mathbf{x}_{d})}\right]\nonumber \\
 &  & +\mathcal{D}_{KL}\left[q(\mathbf{h}_{d}|\mathbf{x}_{d})||p(\mathbf{h}_{d}|\mathbf{y},\mathbf{x}_{d})\right]\nonumber 
\end{eqnarray}
The non-negativity of the Kullback-Leibler divergence $\mathcal{D}_{KL}\left[\cdot\right]$
allows us to only focus on the first term $\mathcal{L}_{B}(\theta,\phi;\mathbf{y},\mathbf{x}_{c},\mathbf{x}_{d})$,
recognizing that this approximation is a lower bound on the overall
true likelihood function, i.e., $\mathcal{L}_{B}(\theta,\phi;\mathbf{y},\mathbf{x}_{c},\mathbf{x}_{d})\leq\mathcal{L}_{A}(\theta,\phi;\mathbf{y},\mathbf{x}_{c},\mathbf{x}_{d})$.

Expanding this approximation term gives us an a separation of log-likelihoods.

\cutequationup
\begin{eqnarray}
 &  & \mathcal{L}_{B}(\theta,\phi;\mathbf{y},\mathbf{x}_{c},\mathbf{x}_{d})\label{eq:approximate-lower-bound-B}\\
 &  & =\int_{\mathbf{h}_{d}}q(\mathbf{h}_{d}|\mathbf{x}_{d})\left[\log\frac{p(\mathbf{y}|\mathbf{x}_{c},\mathbf{h}_{d})p(\mathbf{x}_{d}|\mathbf{h}_{d})p(\mathbf{h}_{d})}{q(\mathbf{h}_{d}|\mathbf{x}_{d})}\right]\nonumber \\
 &  & =\int_{\mathbf{h}_{d}}q(\mathbf{h}_{d}|\mathbf{x}_{d})\left[\log\frac{p(\mathbf{h}_{d})}{q(\mathbf{h}_{d}|\mathbf{x}_{d})}+\log p(\mathbf{y}|\mathbf{x}_{c},\mathbf{h}_{d})+\log p(\mathbf{x}_{d}|\mathbf{h}_{d})\right]\nonumber \\
 &  & =\mathbb{E}_{q(\mathbf{h}_{d}|\mathbf{x}_{d})}\left[\log p(\mathbf{y}|\mathbf{x}_{c},\mathbf{h}_{d})\right]+\mathbb{E}_{q(\mathbf{h}_{d}|\mathbf{x}_{d})}\left[\log p(\mathbf{x}_{d}|\mathbf{h}_{d})\right]\nonumber \\
 &  & -\mathcal{D}_{KL}\left[q(\mathbf{h}_{d}|\mathbf{x}_{d})||p(\mathbf{h}_{d})\right]\nonumber 
\end{eqnarray}
While estimating $q(\mathbf{h}_{d}|\mathbf{x}_{d})$ with sampling
techniques may be possible, a significantly faster method to estimate
this density was a major contribution given by a reparametrization
trick introduced in \cite{kingma2013autoencoding}. In particular,
noting that several common probability densities have scale and location
transformations from ``standard'' random variables, the design feature
representation $\mathbf{h}_{d}\sim q(\mathbf{h}_{d}|\mathbf{x}_{d})$
is instead reparamatrized as a deterministic variable $\mathbf{h}_{d}=g(\mathbf{x}_{d},\mathbf{z}_{d})$,
where $\mathbf{z}_{d}$ is ``standard'' random variable with location
and scale translation properties, and $g_{\phi}(\cdot)$ is a function
now estimated with fast stochastic gradient updates conventional
of deep learning.

Formally, instead of approximating the expectations of the three likelihoods  in Equation (\ref{eq:approximate-lower-bound-B}) using Monte Carlo sampling, below arbitrarily denoted 
$f\left(\cdot,\mathbf{h}_{d}\right)$, 
\begin{eqnarray}
\mathbb{E}_{q\left(\mathbf{\mathbf{h}_{d}}|\mathbf{x}_{d}\right)}\left[f\left(\cdot,\mathbf{h}_{d}\right)\right] & = & \mathbb{E}_{p\left(z\right)}\left[f\left(\cdot,g\left(\mathbf{z}_{d},\mathbf{x}_{d}\right)\right)\right]\label{eq:reparam-1}\\
 & \approx & \frac{1}{L}\sum_{l=1}^{L}f\left(\cdot,g\left(\mathbf{z}_{d}^{(l)},\mathbf{x}_{d}\right)\right)\text{with }\mathbf{z}_{d}^{(l)}\sim p\left(\mathbf{z}_{d}\right)\nonumber 
\end{eqnarray}
in which $s$ denotes Monte Carlo samples and $S$ denotes the total
number of samples, we instead use Equation (\ref{eq:reparam-1})
to reparametrize the lower bound of the full likelihood in Equation
(\ref{eq:approximate-conditional-likelihood}):

\cutequationup 
\begin{eqnarray}
\mathcal{L}\left(\theta,\phi;\mathbf{x}\right) & \approx & \frac{1}{S}\sum_{s=1}^{S}\log\frac{p(\mathbf{h}_{d}^{(s)})}{q(\mathbf{h}_{d}^{(s)}|\mathbf{x}_{d})}+\log p(\mathbf{y}|\mathbf{x}_{c},\mathbf{h}_{d}^{(s)})+\log p(\mathbf{x}_{d}|\mathbf{h}_{d}^{(s)})\nonumber \\
 &  & \text{where }\mathbf{h}_{d}^{(s)}=g\left(\mathbf{z}_{d}^{(l)},\mathbf{x}_{d}\right),\mathbf{z}_{d}^{(l)}\sim q\left(\mathbf{z}_{d}\right)\label{eq:approximate-conditional-likelihood}
\end{eqnarray}
where in practice, this ``sampling'' is one sample to take advantage
of significantly more computationally efficient deep learning methods.

Our second minor extension gives the final probabilistic model we
seek to estimate. Specifically, we define separate members of the
exponential family for design variables; i.e., $p(\mathbf{x}_{d}|\mathbf{h}_{d})$
as as given by their marginal densities in Equation (\ref{eq:bayesian-model-original}).
We follow the original contribution and assume multivariate Gaussian densities,
$q(\mathbf{h}_{d}|\mathbf{x}_{d})$ and $q(\mathbf{z}_{d})$, for
design features.

\cutequationup 
\begin{eqnarray}
p(\mathbf{y}|\mathbf{h}_{c},\mathbf{h}_{d}) & = & \mathrm{M}(\mathbf{y}|\mathbf{h}_{c},\mathbf{h}_{d})\label{eq:overall-deep-combined-model}\\
q\left(\mathbf{h}_{d}|\mathbf{x}_{d}\right) & = & \mathcal{N}\left(\mathbf{h}_{d};\boldsymbol{\mu}\left(\mathbf{x}_{d}\right),\boldsymbol{\sigma}^{2}\left(\mathbf{x}_{d}\right)\right)\nonumber \\
p(\mathbf{x}_{d}|\mathbf{h}_{d}) & = & \mathcal{N}\left(\mathbf{x}_{d,r};\boldsymbol{\mu}\left(\mathbf{h}_{d}\right),\boldsymbol{\sigma}^{2}\left(\mathbf{h}_{d}\right)\mathbf{I}\right)\prod_{b}\mathrm{B}(x_{d,b}|\mathbf{h}_{d})\prod_{c}\mathrm{M}(\mathbf{x}_{d,c}|\mathbf{h}_{d})\nonumber \\
p\left(\mathbf{h}_{d}\right) & = & \mathcal{N}\left(\mathbf{h}_{d};\mathbf{0},\mathbf{I}\right)\nonumber 
\end{eqnarray}
Equation (\ref{eq:overall-deep-combined-model}) details the overall
deep learning model. The first likelihood represents the consumer
choice preference model, while the next three represent the formulation
of the design feasibility model.

This model may be viewed as a deep multinomial logit model over learned
feature embeddings for both consumers $\mathbf{h}_{c}$ and designs
$\mathbf{h}_{d}$, with prior probabilistic structure enforced using
a variational approximation regularizer. In particular, given the
multivariate Gaussian assumption on the approximation distribution
of design features $\mathbf{h}_{d}$, this model may be viewed as
a deep learning generalization of hierarchical Bayes conjoint analysis
\cite{lenk1996hierarchical,allenby1998marketing}. This model may
also be viewed as a deep probabilistic matrix factorization \cite{mnih2007probabilistic}
with incorporation of ``side information,'' with differences being
explicit modeling of the ``side information'' marginal distribution
as well as without probabilistic structure imposed on consumers $\mathbf{h}_{c}$.

These conceptual similarities to existing models lends notion to the
idea that our approach ``learns'' consumer choice preferences both
from the consumer themselves, as well as ``borrowing strength'' from
similar consumers according the learned consumer preference heterogeneity $p(\mathbf{h}_{c})$ and design feasibility ``heterogeneity"
density $p(\mathbf{h}_{d})$.

\subsection{Design Gap Prediction\label{subsec:Design-Gap-Prediction}}

To predict design gaps, the deep learning model of consumer choice
and design feasibility described in Equation (\ref{eq:overall-deep-combined-model})
is used to search for promising regions of the feasible design space.
This search process relies on both an accurate consumer choice model,
as well as an accurate design feasibility model. If either of these
two models has low accuracy, the search process to identify design gaps
will fail.

In addition to accurate consumer choice and design feasibility models,
additional modeling assumptions are still required on the search process
itself and how we combine preferences of several consumers at once.
We accordingly adopt and then describe approaches from design optimization
and information-theoretic measures of disaggregate choice.

At a high-level we draw upon an observation that has been detailed
in various guises from the marketing and design research communities;
namely, that consumer preference functions are \textit{subjective}
and thus fundamentally restricted in their observed assignment of
probability mass due to finite existing designs (or, proposed designs
with stated choice data) \cite{ren2017adaptive,hauser1978testing}.
Even in the perfect case of a fully deterministic utility model for
a given consumer, i.e., Equation (\ref{eq:multinomial-model-original}),
the purchased design may not be the highest possible utility in the
space of all existing and not yet created designs $\mathcal{X}_{d}$.
This is in contrast to \textit{objective} ``preference'' functions
such as object classification models, which in the perfect case, have
``infinite'' utility and assign all of their probability mass to
a single output (e.g., input vehicle is a `BMW').

As a result, we can not simply assume that observed purchases or preference
choices are the ``optimal design\textquotedblright{} for a given
consumer \cite{ren2017adaptive}. Indeed, it would make not make sense
to predict ``design gaps'' if this were the case. Instead, a ``perfect''
subjective preference model over finite existing designs gives us
clues to design needs through its distribution of probability mass.
Therefore, given an appropriately bounded design space, and an accurate
consumer preference model, we can then aim to identify locations of
design gaps.

\cutsubsectionup
\cutsubsectionup
\subsubsection{Information Theory for Disaggregate Choice}

Design gaps are a market-level concept. We are not interested 
in just a single consumer's ``optimal designs,'' but rather regions of
the feasible design space that contain future design concepts that
are preferred by many consumers. Moreover, this should mean either
a large number of consumers who have moderately high preference for
the potential new design concept, or a small number of consumers who
have a very high preference for the potential new design concept. 

To the end, we use $\rho^{2}$ for disaggregate choice to define a
quantitative measure to identify design gaps \cite{hauser1978testing,mokhtarian2016discrete}.

\begin{equation}
\rho^{2}(\mathbf{x}_{d}^{(\hat{d})})=\frac{\Sigma_{c=1}^{C}\delta^{(c,\hat{d})}\log[\frac{p^{*}(y^{(\hat{d})}=1|\mathbf{x}_{d}^{(\hat{d})},\mathbf{x}_{c}^{(c)})}{p^{0}(y^{(\hat{d})}=1|\mathbf{x}_{d}^{(\hat{d})},\mathbf{x}_{c}^{(c)})}]}{\Sigma_{c=1}^{C}\delta^{(c,\hat{d})}\log[\frac{\delta^{(c,\hat{d})}}{p^{0}(y^{(\hat{d})}=1|\mathbf{x}_{d}^{(\hat{d})},\mathbf{x}_{c}^{(c)})}]}
\end{equation}
where $\hat{d}$ is the index of a new proposed design concept, and
$\delta^{(c,\hat{d})}$ is a delta function capturing whether consumer
$c$ purchased design $\hat{d}$. 

This measure describes the explainable amount of ``information''
according to the model relative to a benchmark baseline $p_{cd}^{0}(\mathbf{x}_{d}^{(\hat{d})})$.
This has an information-theoretic interpretation as the entropy reduction
that the consumer choice model achieves normalized by the total amount
of entropy in the data, and as a result $\rho^{2}(\mathbf{x}_{d}^{(\hat{d})})\in[0,1]$.
The denominator is interpretable as a worst case bound on the negative
log-likelihood over all consumers and designs. The delta function
$\delta^{(c,\hat{d})}$ in the logarithm's numerator is a ``perfect''
deterministic model with $0$ entropy throughout the design space
conditioned an arbitrary design from all possible consumers. This
acts as a reference of ``total uncertainty'' for us to compare regions
of the bounded design space using the deep preference model in Equation
(\ref{eq:overall-deep-combined-model}). In our case, $\rho^{2}(\mathbf{x}_{d}^{(\hat{d})})$
can be used to quantitatively identify design gaps by thresholding
the choice model's placement of probability mass throughout the design
space over the set of consumers.

\subsubsection{Design Gap Sampling Algorithm}

To identify possible design gaps, one could optimize $\rho^{2}(\mathbf{x}_{d})$
with respect to $\mathbf{h}_{d}$ over the estimated design space
$p(\mathbf{h}_{d})$. Alternatively, one could sample the space of
designs. In this work, we opted for the latter with a simple rejection
sampling algorithm given in Algorithm \ref{alg:Design-gap-sampling-algorithm}.
Our algorithm is similar in concept to that of sampling the design
space to find ``optimal designs'' as in \cite{ren2017adaptive}.

\begin{algorithm}
\textbf{Input:} $\gamma_{1},\gamma_{2},\hat{D},\rho^{2}(\cdot),p(\mathbf{y}|\mathbf{h}_{c},\mathbf{h}_{d}),q\left(\mathbf{h}_{d}|\mathbf{x}_{d}\right),p(\mathbf{x}_{d}|\mathbf{h}_{d})$

\textbf{Output: }$\left\{ \mathbf{x}_{d}^{(\hat{d})},\rho^{2}(\mathbf{x}_{d}^{(\hat{d})})\right\} _{\hat{d}=1}^{\hat{D}}$

\textbf{Initialize: $p(\mathbf{x}_{d})\approx\hat{p}(\mathbf{x}_{d})=\frac{1}{S}\Sigma_{s=1}^{S}\frac{p_{d}(\mathbf{x}_{d},\mathbf{h}_{d}^{(s)})}{q_{d}(\mathbf{h}_{d}^{(s)}|\mathbf{x}_{d})}$}

\textbf{while:} $\hat{d}\le\hat{D}$ \textbf{do}
\begin{itemize}
\item Sample concept design $\mathbf{x}_{d}^{(\hat{d})}\sim\hat{p}(\mathbf{x}_{d})$ 
\item Reject $\mathbf{x}_{d}^{(\hat{d})}$ if: $-\log[p(\mathbf{x}_{d}^{(\hat{d})})]>\gamma_{1}$ 
\item Calculate $\rho^{2}(\mathbf{x}_{d}^{(\hat{d})})$ for concept design
$\mathbf{x}_{d}^{(\hat{d})}$
\item Reject $\mathbf{x}_{d}^{(\hat{d})}$ if: $\rho^{2}(\mathbf{x}_{d}^{(\hat{d})})<\gamma_{2}$
\item Collect $\mathbf{x}_{d}^{(\hat{d})}$
\end{itemize}
\caption{Design gap sampling algorithm.\label{alg:Design-gap-sampling-algorithm}}
\end{algorithm}

In practice, given the assumption of independence among consumers
$\mathbf{x}_{c}$, we can heuristic sample using a subset of $C_{sub}\le C$
and terminate early if that subset indicates a low likelihood of high
$\rho^{2}(\mathbf{x}_{d}^{(\hat{d})})$ for some empirically assessed
$\rho^{2}$ sampling density $p_{s}$ and threshold $\gamma_{s}$.

\begin{equation}
\mathbb{E}_{p_{s}}\left[\left[-\log[\frac{\Sigma_{c=1}^{C_{sub}}\delta^{(c,\hat{d})}\log[\frac{p^{*}(y^{(\hat{d})}=1|\mathbf{x}_{d}^{(\hat{d})},\mathbf{x}_{c}^{(c)})}{p^{0}(y^{(\hat{d})}=1|\mathbf{x}_{d}^{(\hat{d})},\mathbf{x}_{c}^{(c)})}]}{\Sigma_{c=1}^{C_{sub}}\delta^{(c,\hat{d})}\log[\frac{\delta^{(c,\hat{d})}}{p_{cd}^{0}(y^({\hat{d})}=1|\mathbf{x}_{d}^{(\hat{d})},\mathbf{x}_{c}^{(c)})}]}]>\gamma_{1}\right]\le\gamma_{s}\right]
\end{equation}
Note that all hyperparameters $\gamma_{1},\gamma_{2},\gamma_{s}$
correspond to market-specific thresholds obtained from the ``validation''
set of data.

\section{Experiment\label{sec:Experiment}}

We conduct an experiment to assess how well the proposed deep learning
approach predicts design gaps over the entire U.S. automotive market.
Recall that the utility in this approach is predicting design gaps
\textit{in the future}, with an end goal of early identification of
market opportunities. Validating this approach for its eventual intended
use is currently cost prohibitive, particularly as our focus on the
entire U.S. automotive market would require significant capital investment
and several years of observation.

Instead, we create a scenario with ``artificially induced'' design
gaps in the market. These artificially-induced design gaps are not
simulated, and still uses actual purchase data over real product designs
and real consumers. We retroactively create these design gaps 
using several years of the purchase data by holding out a subset
of these designs, as well as all consumers who purchased those designs
over all years. Next, acting as if we were retroactively looking forward
in time, we aim to identify these design gaps, solely using data from
the ``held-in'' known customers and known designs.

\subsection{Data}

Several proprietary datasets over real purchases of automotive vehicles
in the U.S. market were combined, as well as significant augmentation
of the data by feature engineering by the authors. Feature engineering
refers to manually constructing variables, either from existing data
(e.g., known analytic variable interactions) or from external sources
(e.g., geolocating neighborhood median income).

The combined dataset is truncated to $N=1,000,000$ actual consumer
purchases, and $D=297$ unique product designs of all vehicle types
(i.e., coupes, sedans, SUVs, trucks, vans). Each consumer is represented
using $M_{c}=1574$ variables, and each design is represented using
$M_{d}=2428$ design variables. Moreover, these variables contain
both objective (e.g., `volume of rear seat leg room') and subjective
information (e.g., perceived `sportiness'). Previous research has
show that these types of information must be modeled and processed
differently \cite{burnap2016thesis}.

To preprocess this data, variables are first split by variable types
(i.e., real, binary, categorical) and whether they are objective or
subjective, for a total of 6 possible variable categories. Next, random
indices are generated to split the data into train, validation, and
test sets. The validation dataset is used for selection of model hyperparameters,
model estimation parameters, and design gap sampling hyperparameters.
The test set is only used for calculating final accuracy metrics,
as well as the design gaps themselves.

Depending on the experiment stage, the training, validation, and test
set contain either all $D=297$ designs, or a split of the designs
themselves to induce design gaps. In the latter case, we hold out
30 design as artificially induced design gaps, with the remainder
267 used for learning consumer preferences and the probabilistic encoding
of design constraints. Lastly, we normalize the data according to
the category, using the normalization statistics of the training set
to transform the validation and test set. This process is conducted
using common random seeds for reproducibility.

\subsection{Procedure}

The experiment is divided into three stages: (1) choice model validation,
(2) design constraint model validation, and (3) design gap prediction.
This is due to the notion described in Section \ref{subsec:Design-Gap-Prediction},
in that design gap prediction requires a sufficiently accurate consumer
choice model, as well as a sufficiently accurate design feasibility
model.

Moreover, this split into three stages reflects the three categories
of increasing prediction task challenge described in Section \ref{sec:Introduction}.
In short, the design gap prediction task has the most statistical
unknowns. Instead of predicting held-out customers purchasing held-out
designs, it involves ``not knowing'' the held-out design, and instead
having to search in a high-dimensional design space using the deep
learning approach described by Equation \ref{eq:overall-deep-combined-model}
and the information-theoretic sampler of disaggregate consumer choice
described in Section \ref{subsec:Design-Gap-Prediction}.

To artificially induce design gaps, we randomly hold out 30 vehicles,
15 for validation and 15 vehicles for testing. The testing set induced design gap vehicles
are: Acura RDX, BMW X5, Chevrolet Cobalt, Chrysler PT Cruiser, Ford
Expedition, GMC Acadia Denali, Honda Odyssey, Infiniti M37, Kia Soul,
Lincoln Navigator, Mercury Mariner, Nissan Sentra, Scion tC, Toyota
Corolla, and Volkswagen Jetta, accounting
for a total of 49,497 consumers.

\begin{figure}
\begin{centering}
\includegraphics[width=1\linewidth]{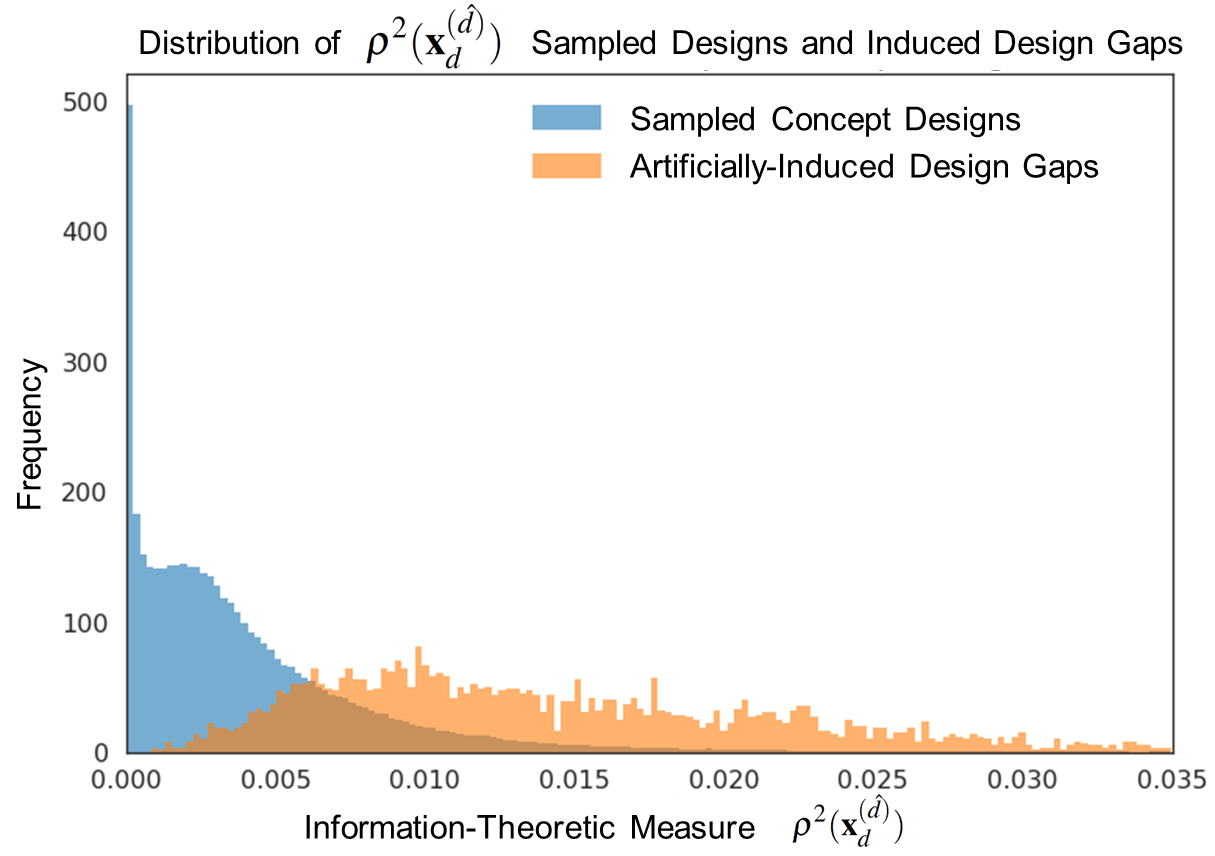}
\par\end{centering}
\caption{Distribution of $\rho^{2}(\mathbf{x}_{d}^{(\hat{d})})$ for the induced
design gaps (orange), and sampled designs $\mathbf{x}_{d}^{(\hat{d})}$.
Note that this histogram is normalized, with many more sampled concepts
that induced design gaps.\label{fig:Distribution-of-rho-squared}}
\end{figure}

\subsubsection{Parameter Estimation}

Maximizing the likelihood of of our model given the data as described
in Equation (\ref{eq:overall-deep-combined-model}) requires balancing
several factors. First, there is an inherent trade off between the
choice model and the desired probabilistic design representation.
Specifically, imposing structure on the design feature representation
$q(\mathbf{h}_{d}|\mathbf{x}_{d})$ through the $\mathcal{D}_{KL}\left[q(\mathbf{h}_{d}|\mathbf{x}_{d})||p(\mathbf{h}_{d})\right]$
term counteracts the desire of the choice model $p(\mathbf{y}|\mathbf{h}_{c},\mathbf{h}_{d})$
to have a complex unknown distribution over $\mathbf{h}_{d}$ in order
to better predict consumer choice. The results in convergence issues
if these seperate terms are not balanced during model estimation.
We accordingly introduce additional model estimation hyperparameters
$\lambda_{1}$ for $\mathcal{D}_{KL}\left[q(\mathbf{h}_{d}|\mathbf{x}_{d})||p(\mathbf{h}_{d})\right]$
within the likelihood optimization objection function.

The architecture of the combined deep learning model is considered
a hyperparameter, and several deep architectures were iterated over
using the training and validation sets. The combined deep learning
model was trained using several first-order optimizers, including
including ADAM \cite{kingma2015adama} and plain stochastic gradient
descent \cite{papalambros2000principles}, as various portions of
the model required different learning rates and parameter freezing.

Due to the heterogeneity of variable types (i.e., real, binary, categorical),
as well as the heteroscadecity in our variables (i.e., vehicle price
range) even after training/validation scaled normalization, several
constraints were imposed during model parameter estimation (e.g.,
minimum estimated variance of $\boldsymbol{\sigma}_{d}^{2}\boldsymbol{I}$).
All experiments were CPU multithreaded and distributed across multiple
GPUs using a single workstation with 128 GB ram, 32 CPU threads at
4.0 GHz, and four Titan XP GPUs.

\begin{table*}[t]
\begin{centering}
\begin{tabular}{c|c|c|c}
\hline 
\textbf{Choice Prediction Task} & \textbf{Top-1 Acc. (std. dev.)} & \textbf{Top-5 Acc. (std. dev.)} & \textbf{Random}\tabularnewline
\hline 
Existing Designs & 83.10\% (0.86\%) & 98.21\% (0.32\%) & 0.34\%\tabularnewline
\hline 
Nonexisting Designs & 76.06\% (0.97\%) & 97.30\% (0.12\%) & 0.37\%\tabularnewline
\hline 
\end{tabular}
\par\end{centering}
\caption{Consumer choice model prediction accuracies using held-out consumers.
Note that ``nonexisting'' designer refers to holding out any data
containing that design (e.g., all Toyota Corollas).
\label{tab:Consumer-choice-model}}
\end{table*}

\subsubsection{Evaluation Metrics}

We aim to predict the exact vehicle the consumer purchases. The consumer
choice model is accordingly assessed using `Top-1' and `Top-5'' evaluation
metrics, where `Top-1' refers to how accurately the choice model predicts
the exact purchased vehicle, and `Top-5' refers to whether the purchased
vehicle is in the choice model's top five predictions. Assuming a
uniform marginal distribution of purchase amongst the designs, the
random chance of correct prediction is either $\frac{1}{297}=0.34\%$,
or $\frac{1}{268}=0.37\%$. Note that in the latter case, while we
hold-in $D=267$, we validate by assuming that only one artificially
induced design gap (i.e., the held-out customer's actual purchase)
enters the market.

To assess the design feasibility model, we use the average negative
log-likelihood (NLL) given by the generative density learned by the
$2^{nd}$ through $4^{th}$ lines in Equation (\ref{eq:overall-deep-combined-model})
on held-out portions of the designs.

Predicting design gaps is not straightforward. We can not, for example,
know whether the design gap sampler given in Algorithm \ref{alg:Design-gap-sampling-algorithm}
is giving false positives (i.e., no design gap), or true positives
that in the past never had a design. Accordingly, we assume that the
only design gaps in the market are those that we can observe using
the artificially induced design gaps from the actual held-out designs. At
the same time, we also assume that every held-out design actually
was a design gap (i.e., no products failed). This assumption invokes
the notion that the U.S. automotive market is relatively mature, and
is dominated by a small set of incumbent players with vast resources,
design intuition, and capability to act on such intuition.

Given these constraints, we use an evaluation metric that assesses
how well the design gap sampler locates designs relative to other
sampled design from the estimated feasible space of designs. Given
the multivariate Gaussian assumption on this space, we use the mean-squared-difference
metric (MSqE).

\begin{table}
\begin{centering}
\begin{tabular}{c|c}
\hline 
\textbf{Design Feasibility Prediction Task} & \textbf{NLL (std. dev.)}\tabularnewline
\hline 
Existing Designs & 2621.88 (-)\tabularnewline
\hline 
Nonexisting Designs & 2220.03 (-)\tabularnewline
\hline 
\end{tabular}
\par\end{centering}
\caption{Negative log likelihood of generative density given design. \label{tab:Design-feasibility-model-accuracies}}
\end{table}

\subsection{Results}

Prediction accuracies for held-out consumers are given in Table \ref{tab:Consumer-choice-model}.
Standard deviations are calculated on 3 experiments with the same
held-out vehicles. Standard deviations are likely larger given more
computational run time for splits with different held-out vehicles.

\begin{table}[t]
\begin{centering}
\begin{tabular}{c|c}
\hline 
\textbf{Design Gap Prediction} & \textbf{MSqE (std. dev.)}\tabularnewline
\hline 
Random Feasible Design & 0.263 (0.000412)\tabularnewline
\hline 
Predicted Gap & 0.251\tabularnewline
\hline 
\end{tabular}
\par\end{centering}
\caption{Mean-squared error between predicted design gaps and randomly sampled
designs.\label{tab:Design-gap-average-distance}}
\end{table}

The evaluation of the design feasibility model is given in Table \ref{tab:Design-feasibility-model-accuracies}.
We note that lower NLL on held-out designs is likely due to the lower
entropy of the these data relative to the learned density. For example,
our random split of testing data did not include designs such as $\frac{1}{2}$-ton trucks or sports convertibles. Further computational runtime with
holding out different random vehicles may lead to different results.

Design gap distances between the artificially induced design gaps,
and the concept designs $\mathbf{x}_{d}^{(\hat{d})}$ sampled by Algorithm
\ref{alg:Design-gap-sampling-algorithm} are given in Table \ref{tab:Design-gap-average-distance}.
Random feasible design refers to any concept design sampled, while
`Predicted Gap' refers to concept designs that were predicted at minimum
to have $\rho^{2}(\mathbf{x}_{d}^{(\hat{d})})\geq\gamma_{2}$ . We
give standard deviations to show the relative scale of this space.
In short, in high-dimensional spaces, all distances looks roughly
the same \cite{aggarwal2001onthe}. Furthermore, in Figure (\ref{fig:Distribution-of-rho-squared}) we plot the distribution
of $\rho^{2}(\cdot)$ for both sampled concept designs $\mathbf{x}_{d}^{(\hat{d})}$
and held-out artificially induced design gaps. 

\section{Discussion and Future Work\label{sec:Discussion}}

While this work is preliminary, the results give evidence that this
approach may have potential for predicting design gaps. This is suggested
by the lower mean-squared distance of sampled design concepts $\mathbf{x}_{d}^{(\hat{d})}$
with $\rho^{2}(\mathbf{x}_{d}^{(\hat{d})})\geq\gamma_{2}$ relative
to those without rejection. This result in itself, however, is not
sufficient to claim prediction of design gaps. These distance calculations
are ultimately being performed in an approximate representation of
a multivariate Gaussian, which is already being estimated using deep
learning models. Deep learning models, while recently state-of-the-art
for many prediction tasks, are known to be ``brittle'' in their predictions
\cite{goodfellow2014explaining}. This can lead to warping of the
design space that lead to unreasonable distance calculations, particularly
given our prediction regime of held-out customers, held-out designs,
and unknown design gaps.

At the same time, these results in conjunction with the very strong
prediction results by the consumer choice model given in Table \ref{tab:Consumer-choice-model}
do suggest that this approach may have have potential for more direct
estimation of design gaps. Given the implications of a validated method
to help product design and market opportunities potentially years
in advance, without requiring \textit{a priori} knowledge possible
design concepts, further study is warranted.

Accordingly, there are significant opportunities for future work.
Perhaps the portion of this approach with most opportunity for improvement
lies in the search method of the bounded design space. The sampling
method given in Algorithm \ref{alg:Design-gap-sampling-algorithm},
like many sampling algorithms, is computationally inefficient due
to the high-dimensionality space the samples are drawn samples (even
after dimensionality reduction from the original design variable representation).
With better validation on the structure of the feature space, optimization
methods would likely prove to be a significantly more computationally
efficient means of searching the space of possible designs for design
gaps.

This opportunity to perform optimization over this space perhaps deserves
attention on its own. In short, this is enabled by the magnitude of
prediction accuracy obtained by the deep consumer choice model. This
work builds off previously conducted work by the authors also in vehicle
purchase prediction \cite{burnap2016improving}, in which the task
was of more conventional binary choice conjoint analysis. We note
that in the previous study, an accuracy of $75.15\%$ was achieved
by the model most conceptually similar to that in this work. However,
in the previous study $50\%$ represented random chance while in this
study, random chance is effectively $0\%$. These two tasks are best
conceptually comparable if we were to decompose each multinomial prediction
to all pairwise comparisons. A consumer choice model that has less
than 50\% accuracy on average for all pairwise design comparisons
will always lead to suboptimal design.

As earlier noted with regards to distance metrics, efficiently searching
the design space using optimization also requires well-behaved probabilistic
structure imposed on the design feature space. This comes with a cost
of (potentially significantly) reducing the predictive accuracy of
the consumer choice model. This tradeoff thus warrants further study.
Lastly, any future improvement and even development of this approach
should necessitate ``real'' validation for actual usage. In other
words, without using data alone, controlled experiments should be
conducted to with new consumers to assess to validity of this approach.

\section{Conclusion\label{sec:Conclusion}}

In this work we introduce a deep learning approach to predict design
gaps by learning a consumer choice model as well as a design feasibility
model. This approach builds on conventions in both quantitative
marketing in bounding the heterogeneity of consumer choice preferences, as well as engineering design
for bounding the space of possible designs.
We further introduce variational approximations to induce desired
probabilistic structure in the space of possible designs, as well
as making this model amenable for efficient parameter estimation using
recent advances in machine learning. Our approach is tested on a large
dataset of real design purchases in the U.S. automotive market. While
our work is preliminary, we find that evidence that with sufficiently
accurate consumer choice and design constraint models, it may be possible
to predict design gaps in the market that do not need to be specified
beforehand.

\section*{Acknowledgment}

This research was partially supported by grants from Pangaea Research, NVIDIA Corporation and several
data providers including DataOne Software. This support is gratefully
acknowledged.

\bibliographystyle{plain}
\bibliography{idetc2018_design_gap}

\end{document}